\documentclass[11pt]{article}
\usepackage{amsfonts}
\usepackage{lscape}
\usepackage{latexsym,amsmath}
\usepackage{amssymb,array}
\makeatletter 
\RequirePackage[dvips]{graphicx} 
\newtheorem{t1}{Theorem}[section]

\newtheorem{l1}{Lemma}[section]

\begin{document}
\title{\textbf{Some estimators of the PDF and CDF of the Lindley Distribution}}
\author{Sudhansu S. Maiti\footnote{Corresponding author. e-mail:
dssm1@rediffmail.com} and Indrani Mukherjee\\
Department of Statistics, Visva-Bharati
University, Santiniketan-731 235, West Bengal, India}
\date{}
\maketitle
\begin{center}
Abstract
\end{center}
This article addresses the different methods of estimation of the probability density function (PDF) and the cumulative distribution function (CDF) for the Lindley distribution. Following estimation methods are considered: uniformly minimum variance unbiased estimator (UMVUE), maximum likelihood estimator (MLE), percentile estimator (PCE), least square estimator (LSE), weighted least square estimator (WLSE), Cram\'{e}r-von-Mises estimator (CVME), Anderson-Darling estimator (ADE). Monte Carlo simulations are performed to compare the performances of the proposed methods of estimation.

\vspace{0.5cm}

\noindent \textbf{Keywords:}  Maximum likelihood estimators; percentile estimators; uniformly minimum variance unbiased estimators; least square estimators; weighted least square estimators; Cram\'{e}r-von-Mises estimators; Anderson-Darling estimators.
\\ {\bf 2010 Mathematics Subject Classification.} 62E15, 62N05, 62P30, 
\newpage
\section{Introduction}
\ A random variable $X$ is said to have the Lindley distribution, if its probability density function (PDF) is given by
\begin{eqnarray}\label{lindpdf}
f(x)=\frac{\theta ^2}{1+\theta}(1+x)e^{-\theta x},~ x,~ \theta >0
\end{eqnarray}
and its cumulative distribution function (CDF) is given by
\begin{eqnarray}
F(x)=1-\frac{1+\theta +\theta x}{1+\theta} e^{-\theta x},~ x>0,~ \theta >0.
\end{eqnarray}
The above distribution is close to the exponential distribution. But many of the mathematical properties (e.g., the mode of the distribution, moments, skewness and kurtosis measures, cumulants, failure rate and mean residual life, mean deviation, entropies etc.) are more flexible than those of exponential distribution.
\par Now a days researchers have given attention for study of properties and inference on this distribution. Some extension models have been found out and their properties and statistical inferences are made by host of authors. Statisticians are most of the times interested about inferring the parameter(s) involved in the distribution. MLE and Bayes estimate of the parameter has been focused by the authors. Hardly any unbiased estimator of the parameter has been studied so far and finding out MVUE of the parameter seems to be intractable and consequently the comparison with any unbiased class of estimator is not being made. However instead of studying the estimators of the parameter(s), we have scope to find out unbiased estimator of the PDF and the CDF as well as some biases estimator of the same and comparison among the estimators could be made. That is why we have shifted our focus from estimation of parameter(s) to estimation of the PDF and the CDF.
\par We see many situations where we have to estimate PDF, CDF or both. For instance, PDF can be used for estimation of differential entropy, R\'{e}nyi entropy, Kullback-Leibler divergence and Fisher information; CDF can be used for estimation of cumulative residual entropy, the quantile function, Bonferroni curve, Lorenz curve, and both PDF and CDF can be used for estimation of probability weighted moments, hazard rate function, mean deviation about mean etc.
\par Several papers are available over this similar type of studies. As for example
Asrabadi [$\ref{Asrabadi-1990}$], Dixit and Jabbari [$\ref{Dixit-Nooghabi-2010}$], Dixit and Jabbari [$\ref{Dixit-Jabbari-2011}$], Jabbari and Jabbari [$\ref{Jabbari- Jabbari-2010}$], Bagheri et al. [$\ref{bagheri}$], Bagheri et al. [$\ref{Bagheri-Alizadeh-Nadarajah-2016}$], Bagheri et al. [$\ref{Bagheri-Alizadeh-Jamkhaneh-Nadarajah-2013}$], Alizadeh et al. [$\ref{alizadeh2015}$] and the references cited therein.
\section{Maximum likelihood estimators of the PDF and the CDF}
\ Let $X_1, X_2,...,X_n$ be a random sample of size n with PDF $\eqref{lindpdf}$. The MLE of $\theta$ say $\tilde{\theta}$ is
\begin{equation}
\widetilde{\theta}=\frac{-(\overline{x}-1)+\sqrt{(\overline{x}-1)^2+8\overline{x}}}{2\overline{x}}
\end{equation}
\\Therefore, by using the invariance property of MLE, one can obtain the MLEs of the PDF and the CDF as
\begin{equation}
\widetilde{f}(x)=\frac{\widetilde{\theta} ^2}{1+\widetilde{\theta}}(1+x)e^{-\widetilde{\theta} x};~ x>0,~\theta >0
\end{equation}
and
\begin{equation}
\widetilde{F}(x)=1-\frac{1+\widetilde{\theta} +\widetilde{\theta} x}{1+\widetilde{\theta}} e^{-\widetilde{\theta} x};~ x>0,~ \theta >0
\end{equation}
respectively.

Here $T=t$ is a complete sufficient statistic. If we replace $\overline{x}$ by $\frac{t}{n}$ then we get 
\begin{equation}\label{gt}
g(t)=\frac{-(t-n)+\sqrt{(t-n)^2+8tn}}{2t}=\widetilde{\theta}
\end{equation}
\begin{t1}\label{ft}
(Ghitany et al.[$\ref{Ghitany}$]) If $X_1,~X_2,~\ldots,~X_n$ are IID RVs from Lindley ($\theta$), then the PDF of $T=X_1+X_2+\ldots +X_n$ is 
\begin{equation}
f(t;n,\theta)=\sum _{k=0}^n p_{k,n}(\theta)f_{GA}(t;2n-k,\theta)~,
\end{equation}
where $p_{k,n}=\binom {n}k \frac{\theta ^k}{(1+\theta)^n}$ and $f_{GA}(t;m,\theta)=\frac{\theta ^m}{\Gamma(m)}t^{m-1}
e^{-\theta t}$, $t>0$, is the PDF of gamma distribution with shape and scale parameters $m$ and $\theta$, respectively.
\end{t1}

\vspace{0.2in}

\begin{t1}\label{th1}
The estimators, $\widetilde{f}(x)$ and $\widetilde{F}(x)$, are biased for $f(x)$ and $F(x)$, respectively, with
\begin{equation*}
E(\widetilde{f}(x))=\int _0^\infty \frac{g(t)^2}{1+g(t)}(1+x)e^{-g(t)x}f(t)|\frac{dr}{dt}|dt
\end{equation*}
and
\[
E(\widetilde{F}(x))=\int _0^\infty \left[1-\frac{g(t)+1+g(t)x}{1+g(t)}e^{-g(t)x}\right]f(t)|\frac{dr}{dt}|dt
\]
\end{t1}
where the value of $\frac{dr}{dt}$ is
\begin{equation*}
\frac{dr}{dt}= \left[-\frac{n}{2t^2}+\frac{t+3n}{2t\sqrt{(t-n)^2+8tn}}-\frac{\sqrt{(t-n)^2+8tn}}{2t^2}\right].
\end{equation*}

\vspace{0.2in}

\begin{t1}\label{th2}
The MSEs of $\widetilde{f}(x)$ and $\widetilde{F}(x)$ are given by

\vspace{0.2in}

$MSE(\widetilde{f}(x))=\int _0^\infty \left[\frac{g(t)^2}{1+g(t)}(1+x)e^{-g(t)x}-f(x)\right]^2 f(t)|\frac{dr}{dt}|dt$

\vspace{0.2in}

and

\vspace{0.2in}

$MSE(\widetilde{F}(x))=\int _0^\infty \left[1-\frac{g(t)+1+g(t)x}{1+g(t)}e^{-g(t)x}-F(x)\right]^2 f(t)|\frac{dr}{dt}|dt$
\end{t1}

\section{\textbf{UMVU estimators of the PDF and the CDF}}
\ In this section, we obtain the UMVU estimators of the PDF and the CDF of the Lindley distribution. Also, we obtain the MSEs of these estimators.

\vspace{0.2in}

\begin{l1}\label{umvue}
(Ghitany et al.[$\ref{Ghitany}$]) If $X_1,~X_2,~\ldots,~X_n$ be n IID Lindley ($\theta$), then the conditional PDF of $X_1$ given $T=\sum_{i=1}^n X_i$ is 

\vspace{0.2in}
\begin{equation*}
f_{X_1|T=t}(x)=\frac{1+x}{A_n(t)}\sum_{k=0}^{(n-1)}C_{k,n}(t-x)^{2n-3-k},~~0~<~x~<~t
\end{equation*}
where $C_{k,n}=\frac{\binom {n-1}k}{\Gamma(2n-2-k)}$ and ${A_n(t)}=\sum_{j=0}^n \binom nj \frac{t^{2n-j-1}}{\Gamma(2n-j)}$.
\end{l1}

\vspace{0.2in}

\begin{t1}\label{umvuefandF}
Let $T=t$ be given. Then
\begin{equation*}
\widehat{f}(x)=\frac{1+x}{A_n(t)}\sum_{k=0}^{(n-1)}C_{k,n}(t-x)^{2n-3-k},~~0~<~x~<~t
\end{equation*}
is an UMVUE for $f(x)$ and
\begin{equation*}
\widehat{F}(x)=\frac{1}{A_n(t)}\sum_{k=0}^{n-1}C_{k,n}t^{2n-2-k}\left[\frac{1}{2n-k-2}I_{x/t}(1,2n-2-k)
+t\frac{1}{(2n-k-1)(2n-k-2)}I_{x/t}(2,2n-2-k)\right],~~0~<~x~<~t
\end{equation*}
is an UMVUE for $F(x)$.
Here $I_{x}(\alpha ,\beta)=\frac{1}{B(\alpha ,\beta)}\int_0^x x^{\alpha -1}(1-x)^{\beta -1}$ is an incomplete beta function and 
$B(\alpha ,\beta)=\frac{\Gamma \alpha\Gamma \beta}{\Gamma (\alpha +\beta)}$.
\end{t1}

\vspace{0.2in}

\begin{t1}

\vspace{0.2in}

The MSE of $\widehat{f}(x)$ is given by
\begin{equation*}
MSE(\widehat{f}(x))=\int _x^\infty \left[\frac{1+x}{A_n(t)}\sum_{k=0}^{(n-1)}C_{k,n}(t-x)^{2n-3-k}\right]^2f(t)dt-f^2(x)
\end{equation*}
\ and
\begin{eqnarray*}
MSE(\widehat{F}(x))&=&\int _x^\infty [\frac{1}{A_n(t)}\sum_{k=0}^{n-1}C_{k,n}(t^2\frac{1}{2n-k-2}I_{x/t}(1,2n-2-k)\\
&&+t^3\frac{1}{(2n-k-1)(2n-k-2)}I_{x/t}(2,2n-2-k))]^2f(t)dt-F^2(x)
\end{eqnarray*}

\end{t1}

\vspace{0.2in}

\section{\textbf{Least squares and weighted least squares estimators}}
\ The least square estimators and weighted least square estimators were proposed by Swain et al. [\ref{Sw88}] to estimate the parameters of Beta distributions. In this paper, we apply the same technique for the Lindley distribution. Suppose $X_1,..., X_n$ is a random sample of size $n$ from a CDF $F(.)$ and let $X_{i:n}$, $i=1,...,n$ denote the ordered sample in ascending order. The proposed method uses the CDF of $F(x_{i:n})$. For a sample of size $n$, we have $E[F(X_{j:n})]=\frac{j}{n+1}$, $Var[F(X_{j:n})]=\frac{j(n-j+1)}{(n+1)^2(n+2)}$ and $Cov[F(X_{j:n}),F(X_{k:n})]=\frac{j(n-k+1)}{(n+1)^2(n+2)}$ for $j<k$, (see Johnson et al. [$\ref{Johnson-Kotz-Balakrishnan-1994}$]). Using the expectations and the variances, two variants of the least squares method follow.

\vspace{0.2in}

\textbf{Method 1: Least squares estimators}

\vspace{0.2in}

This method is based on minimizing   

\vspace{0.2in}

$\sum_{j=1}^n [F(X_{j:n})-\frac{j}{n+1}]^2$

\vspace{0.2in}

with respect to the unknown parameters. 

\vspace{0.2in}
\begin{itemize}
\item In case of \textbf{Lindley Distribution} the least squares estimators of $\theta$ is $\widetilde{\theta}_{LSE}$.
$\widetilde{\theta}_{LSE}$ can be obtained by minimizing

\vspace{0.2in}

$\sum_{j=1}^n[1-\frac{1+\theta +\theta x_{j:n}}{1+\theta} e^{-\theta x_{j:n}}-\frac{j}{n+1}]^2$
with respect to $\theta$.

\vspace{0.2in}

So, to obtain the  LS estimators of the PDF and the CDF, we use the same method as for the MLE. Therefore,
\begin{equation}
\widetilde{f}_{LSE}(x)=\frac{\widetilde{\theta}_{LSE} ^2}{1+\widetilde{\theta}_{LSE}}(1+x)e^{-\widetilde{\theta}_{LSE}~ x}
\end{equation}
and
\begin{equation}
\widetilde{F}_{LSE}(x)=1-\frac{1+\widetilde{\theta}_{LSE} +\widetilde{\theta}_{LSE}~ x}{1+\widetilde{\theta}_{LSE}} 
e^{-\widetilde{\theta}_{LSE}~ x}
\end{equation}
It is difficult to find the expectations and the MSE of these estimators analytically, so we calculate them by means of simulation study.
\end{itemize}

\vspace{0.2in}

\textbf{Method 2: Weighted Least squares estimators}

\vspace{0.2in}

This method is based on minimizing

\vspace{0.2in}

$\sum_{j=1}^n w_j[F(X_{j:n})-\frac{j}{n+1}]^2$

\vspace{0.2in}

with respect to the unknown parameters, where

\vspace{0.2in}

$ w_j=\frac{1}{Var[F(X_{j:n})]}=\frac{(n+1)^2(n+2)}{j(n-j+1)}$

\vspace{0.2in}

\begin{itemize}
\item In case of the \textbf{Lindley distribution}, the weighted least squares estimators of $\theta$ say 
$\widetilde{\theta}_{WLSE}$ is the value minimizing 

\vspace{0.2in}

$\sum_{j=1}^n w_j[1-\frac{1+\theta +\theta x_{j:n}}{1+\theta} e^{-\theta x_{j:n}}-\frac{j}{n+1}]^2$.

\vspace{0.2in}

So,  the WLS estimators of the PDF and CDF are
\begin{equation}
\widetilde{f}_{WLSE}(x)=\frac{\widetilde{\theta}_{WLSE} ^2}{1+\widetilde{\theta}_{WLSE}}(1+x)e^{-\widetilde{\theta}_{WLSE}~ x}
\end{equation}
and
\begin{equation}
\widetilde{F}_{WLSE}(x)=1-\frac{1+\widetilde{\theta}_{WLSE} +\widetilde{\theta}_{WLSE}~ x}{1+\widetilde{\theta}_{WLSE}} 
e^{-\widetilde{\theta}_{WLSE}~ x}
\end{equation}
It is difficult to find the expectations and the MSE of these estimators analytically. So, we can calculate them by means of a simulation study. 
\end{itemize}

\section{\textbf{Estimators based on percentiles}}
\ Estimations based on percentiles was originally  suggested by Kao [$\ref{Ka58}$, $\ref{Ka59}$].
Percentiles estimators are based on inverting the CDF. Since the Lindley distribution has a closed form CDF, its parameters can be estimated using percentiles.
\begin{itemize}
\item Let $X_{i:n}$, $i=1,...,n$ denote the ordered random sample from the Lindley distribution. Also let $p_i=\frac{i}{n+1}$.
The percentile estimator of $\theta$ say $\widetilde{\theta}_{PCE}$ is the value minimizing

\vspace{0.2in}

$\sum_{i=1}^n [ log(1-p_i)-log\left(\frac{1+\theta+\theta x_{i:n}}{1+\theta}\right)-\theta x_{i:n}]^2$.

\vspace{0.2in}

So, the percentile estimators of the PDF and CDF are
\begin{equation}
\widetilde{f}_{WLSE}(x)=\frac{\widetilde{\theta}_{PCE} ^2}{1+\widetilde{\theta}_{PCE}}(1+x)e^{-\widetilde{\theta}_{PCE}~ x}
\end{equation}
and
\begin{equation}
\widetilde{F}_{WLSE}(x)=1-\frac{1+\widetilde{\theta}_{PCE} +\widetilde{\theta}_{PCE}~ x}{1+\widetilde{\theta}_{PCE}} 
e^{-\widetilde{\theta}_{PCE}~x}
\end{equation}
The expectations and the MSE of these estimators can be calculated by simulation.

\end{itemize}

\section{Estimator based on  Cram\'{e}r-von-Mises }
\ To motivate our choice of Cram\'{e}r-von-Mises  type minimum distance estimators, MacDonald [$\ref{Ma71}$] provided empirical evidence that the bias of the estimator is smaller than the other minimum distance estimators. Thus,
the Cram\'{e}r-von-Mises  estimator $\widetilde{\theta}_{CVME}$  of the parameter $\theta $ is
obtained by minimizing, with respect to $\theta $  the function:
\begin{equation}
C(\theta) =\frac{1}{12n}+\sum_{i=1}^{n}\left( F\left(
x_{i:n}\mid \theta\right) -{\frac{2i-1}{2n}}\right) ^{2}.
\end{equation}
The estimator can also be obtained by solving the following non-linear
equation:
\begin{eqnarray*}
\sum_{i=1}^{n}\left( F\left( x_{i:n}\mid \theta \right) -{\frac{2i-1}{2n%
}}\right) \Delta \left( x_{i:n}\mid \theta \right)  &=&0, 
\end{eqnarray*}
where $\Delta \left( . \mid \theta \right) $  is given by 
\begin{equation}\label{delta1}
\Delta \left( x_{i:n}\mid \theta \right) = \frac{(x_{i:n}^2\theta+x_{i:n}\theta -1)e^{-\theta x_{i:n}}}{1+\theta}
+\frac{(1+\theta +\theta x_{i:n})e^{-\theta x_{i:n}}}{(1+\theta)^2}.
\end{equation}

\section{Estimator based on Anderson-Darling }
\ The Anderson-Darling test was developed in 1952 by T.W. Anderson and D.A.Darling [$\ref{AnDa52}$] as an alternative to other statistical tests for detecting sample distributions departure from normality. Specifically,  the AD test  converge very quickly towards the asymptote (Anderson \& Darling [$\ref{AnDa54}$]; Pettitt [$\ref{Pe76}$]; Stephens [$\ref{St74}$]).
The Anderson-Darling estimator $\widetilde{\theta}_{ADE}$  of the parameter $\theta$ is obtained
by minimizing, with respect to $\theta$ , the function:%
\begin{equation}
A(\theta) =-n-\frac{1}{n}\sum_{i=1}^{n}\left( 2i-1\right)
\left\{ \log F\left( x_{i:n}\mid \theta \right)+ \log \overline{F}\left(
x_{n+1-i:n}\mid \theta \right)\right\} .
\end{equation}
The estimator can also be obtained by solving the following non-linear equation:
\begin{eqnarray*}
\sum_{i=1}^{n}\left( 2i-1\right) 
\left[ \frac{\Delta\left( x_{i:n}\mid \theta \right) }{F\left( x_{i:n}\mid \theta \right) }-
\frac{\Delta \left( x_{_{n+1-i:n}}\mid \theta \right) }
{\overline{F}\left( x_{n+1-i:n}\mid \theta \right) }\right] &=&0, 
\end{eqnarray*}
where $\Delta \left( \cdot \mid \theta \right) $  is given by $\left( \ref{delta1} \right) $ .

\section{Simulation study}
\ Here, we conduct Monte Carlo simulation to evaluate the performance of the estimators for the PDF and the CDF discussed in the previous sections. All computations were performed using the R-software. We evaluate the performance of the estimators based on MSEs. The MSEs were computed by generating $1000$ replications from Lindley Distribution. It is observed that MSEs decreases with increasing sample size. It verifies the consistency properties of all the estimators. We observe from true MSE point of view, MLE is better than UMVUE for both PDF and CDF.

\section{ Conclusion}
In this article, different methods of estimation of the probability density function and the cumulative distribution function of  the Lindley distribution have been considered. Uniformly minimum variance unbiased estimator (UMVUE), maximum likelihood estimator (MLE), percentile estimator (PCE), least square estimator (LSE) and weighted least square estimator (WLSE), Cram\'{e}r-von-Mises estimator (CVME), Anderson-Darling estimator (ADE) have been found out. Monte Carlo simulations are performed to compare the performances of the proposed methods of estimation. If we restrict to unbiased class of estimators, UMVUE is better in minimum variance sense. Though MLEs are in biased class, it is preferable in MSE sense. 

\end{document}